\documentclass[pra,showpacs,twocolumn]{revtex4-1}
\usepackage{}
\usepackage{amsfonts}
\usepackage{amssymb}
\usepackage{amsmath}
\usepackage{graphicx}
\usepackage{savesym}
\savesymbol{iint}
\usepackage{txfonts}
\restoresymbol{TXF}{iint}
\usepackage{mathrsfs}
\usepackage{bbm}
\usepackage{bm}
\usepackage{graphicx}

\newcommand{\ket}[1]{\left\vert#1\right\rangle}
\newcommand{\bra}[1]{\left\langle#1\right\vert}

\begin{document}

\title{Using Nonlocal Coherence to Quantify Quantum Correlation}
\author{Pei Pei}
\author{Wei Wang}
\author{Chong Li}
\email{lichong@dlut.edu.cn}
\author{He-Shan Song}
\affiliation{School of Physics and Optoelectronic Engineering, Dalian University of Technology, Dalian 116024, P. R. China}
\date{\today}
\begin{abstract}
We reexamine quantum correlation from the fundamental perspective of its consanguineous quantum property, the coherence. We emphasize the importance of specifying the tensor product structure of the total state space before discussing quantum correlation. A measure of quantum correlation for arbitrary dimension bipartite states using nonlocal coherence is proposed, and it can be easily generalized to the multipartite case. The quantification of non-entangled component within quantum correlation is investigated for certain states.
\end{abstract}
\maketitle

\begin{figure}
\includegraphics[width=4.5cm]{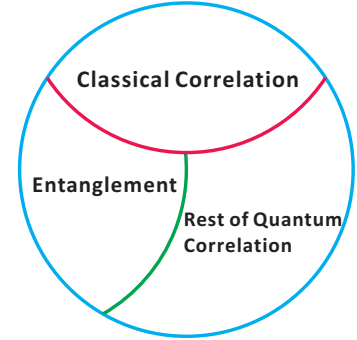}
\caption{\label{fig:1}(Color online) Schematic diagram for the components of correlation. The red curve corresponds to the boundary between classical and quantum correlations, while the green curve to the boundary between entanglement and non-entangled quantum correlation.}
\end{figure}

The correlation between generic quantum systems is divided into two kinds: classical and quantum. The former can be prepared with the local operations and classical communication (LOCC) while the latter cannot. It is a crucial and basic problem to characterize and quantify these two kinds of correlations in quantum information theory. To address this issue, \textit{entanglement} \cite{Einstein35,*Schrodinger35} is in widespread use as one aspect of quantum correlation, and abundant criterions and measures \cite{Horodecki09} centering around the entanglement vs separability paradigm scenario \cite{Werner89} have been invested. However, entanglement only quantifies part of quantum correlation (see Fig.~\ref{fig:1}) since there exist separable states also showing nonclassicality, for example the already demonstrated ``quantum nonlocality without entanglement'' \cite{Bennett99,Horodecki05,Niset06,Braunstein99,Meyer00,Biham04,Datta05,*Datta07}. Therefore, other measures of quantum correlations are expedited subsequently to ascertain the ``dividing line'' (red curve in Fig.~\ref{fig:1}) between classical and quantum correlations more exactly from various perspectives, e.g. the quantum extensions of the classical mutual information \cite{Henderson01,Ollivier01}, the connections between thermodynamics and information \cite{Oppenheim02}, the work to erase the correlation \cite{Groisman05}, and the disturbance induced by generic measurements \cite{Luo08a}. These studies concern mainly bipartite correlations, and it is desirable to distinguish bipartite and multipartite correlations among multipartite systems. One may think investigating quantum correlations from another fundamental view will lead to a measure not only to find the ``dividing line'' for bipartite systems, but also easy to be generalized to only quantify genuine multipartite quantum correlation for multipartite systems.


On the other hand, most of previous works are focused on dichotomizing the total correlation without concerning the components within the quantum correlation. Hence an interesting question arises as to how to characterize and quantify the rest of quantum correlation excluding entanglement, namely how to ascertain the ``dividing line'' (green curve in Fig.~\ref{fig:1}) between entanglement and non-entangled quantum correlation. A simple subtraction is infeasible due to the quantitatively and qualitatively difference between the notions of quantum correlation and entanglement, for instance the incomparability between quantum discord and the entanglement of formation \cite{Luo08b}. Recently, Modi \textit{et al}. \cite{Modi10} redefine the measures of quantum and classical correlations under a unified frame based on the geometric
measure of nearest distance between the given state and the state without the desired property, and propose quantum \textit{dissonance} to quantify the rest of nonclassical correlation. But it appears ambiguous that whether dissonance can ascertain the ``dividing line'' clearly. In this paper, we firstly reexamine the quantum correlation from the perspective of its consanguineous quantum property, the quantum \textit{coherence}, then propose a measure to quantify quantum correlation for arbitrary dimension bipartite states using nonlocal coherence. For certain states the non-entangled quantum correlation is investigated considering the qualitatively similar measure of entanglement.

Quantum correlation and coherence both arise from quantum \textit{superposition}. Consider a quantum compound system in a coherent superposition state, if there exists coherence with respect to the basis of a certain product Hilbert space, the nonlocal coherence may induce the nonclassical correlation (in a narrow sense, entanglement) between the corresponding subsystems, besides the local coherence of each corresponding subsystem \cite{Yu09}. In this case, quantum correlation is a kind of nonlocal coherence. Note that usually when we talk about quantum correlation and coherence, a standard structure of the Hilbert space of the given compound system has been determined beforehand. However Zanardi \textit{et al}. \cite{Zanardi04} demonstrated that whether a state is entangled depends on the tensor product structure (TPS) of the state space. Therefore before proposing our measure, we recall their points with explicit examples and emphasize that in a general sense, the specification of TPS of the total Hilbert space is necessary not only for entanglement, but also for other quantum correlations.

For entangled states, taking the state $\ket{\phi}_{12}=a_1\ket{00}_{12}+a_2\ket{11}_{12}$ and $\ket{\phi}_{34}=b_1\ket{00}_{34}+b_2\ket{11}_{34}$ for example, there is no
entanglement between the subsystems with Hilbert spaces $H_{12}$ and $H_{34}$, but
there exists entanglement between the subsystems with Hilbert spaces $H_{13}$ and $H_{24}$,
\begin{eqnarray}
\ket{\Phi}_{1234}&=&\ket{\phi}_{12}\otimes\ket{\phi}_{34}\nonumber\\
&=&a_{1}b_{1}\ket{00}_{13}\ket{00}_{24}+a_{1}b_{2}\ket{01}_{13}\ket{01}_{24}\nonumber\\
&+&a_{2}b_{1}\ket{10}_{13}\ket{10}_{24}+a_{2}b_{2}\ket{11}_{13}\ket{11}_{24}.
\end{eqnarray}


Besides the entanglement, whether other quantum correlations exists also depends on the specification of TPS of the state space, even with the absence of entanglement. Consider the Werner state \cite{Werner89}, $\rho=a\ket{\Psi^-}\bra{\Psi^-}+(1-a)I/4$, where $\ket{\Psi^-}=(\ket{0}_1\ket{1}_2-\ket{1}_1\ket{0}_2)/\sqrt{2}$ is the maximally entangled state and $0\leq{a}\leq1$. It is well known that the concurrence \cite{Wooters98} for the Werner state is given as $C(\rho)=\max\{0,(3a-1)/2\}$, and the quantum discord as $\mathcal{D}(\rho)=(1/4)[(1-a)\log_2(1-a)+(1+3a)\log_2(1+3a)-2(1+a)\log_2(1+a)]$ \cite{Luo08b,Ali10a}. For $0<{a}\leq1/3$ the entanglement is vanishing while the quantum discord is still positive. The acquiescent TPS of the state space is $H={H_{1}}\otimes{H_{2}}$, and Bell states $\left\{\ket{\Psi^\pm},\ket{\Phi^\pm}\right\}$ is a group of orthogonal basis of space $H$. If one redefine the original basis as $\ket{\Psi^\pm}=\ket{\Psi}_1\otimes\ket{\pm}_2$ and $\ket{\Phi^\pm}=\ket{\Phi}_1\otimes\ket{\pm}_2$ \cite{Zanardi04}, thus the TPS is mapped as
\begin{eqnarray}
F':H={H_{1}}\otimes{H_{2}}\longrightarrow{\tilde{H}_{1}}\otimes{\tilde{H}_{2}},
\end{eqnarray}
where $\tilde{H}_{1}=\{\ket{\Psi}_{1},\ket{\Phi}_{1}\}$ and $\tilde{H}_{2}=\{\ket{+}_{2},\ket{-}_2\}$. With the new TPS the Werner state is rewritten as
\begin{eqnarray}
\rho=a\ket{\Psi}_1\ket{-}_2\bra{\Psi}_1\bra{-}_2+\frac{1-a}{4}\tilde{I},
\end{eqnarray}
where $\tilde{I}$ is the identity operator with the basis $\{\ket{\Psi}_1\ket{+}_2,\ket{\Psi}_1\ket{-}_2,\ket{\Phi}_1\ket{+}_2,\ket{\Phi}_1\ket{-}_2\}$, then $\rho$ is now a classical state and discord is vanishing even for $0\leq a\leq 1/3$.

Now we specify the TPS of the total Hilbert space for arbitrary dimension bipartite quantum states. Let set $\{\ket{k}_1|k=1,2,\cdots,d_1\}$ and $\{\ket{k}_2|k=1,2,\cdots,d_2\}$ be the group basis of $d$-dimension Hilbert spaces $H^{(1)}$ and $H^{(2)}$, respectively. Then the TPS of the total Hilbert space is $\mathcal{C}=H^{(1)}\otimes H^{(2)}$. For pure states, any state $\ket{\phi}$ can be written with the product basis as $\ket{\phi}=\sum_{k_{1}k_{2}}c_{k_{1}k_{2}}\ket{k_1}_1\otimes\ket{k_2}_2$,
where $c_{k_{1}k_{2}}$ is a complex number. We perform proper local unitary operation $U_{1}$ and $U_2$ on state $\ket{\phi}$, if we obtain $U_{1}\otimes U_{2}\ket{\phi}=\sum_kP_{k}\ket{\tilde{k}}_{1}\otimes\ket{\tilde{k}}_{2}$,
then we can say state $\ket{\phi}$ has been Schmidt decomposed. If any $|P_{k}|\neq1$, state $\ket{\phi}$ is an entangled state, namely there exists quantum correlation between subsystems with Hilbert spaces $H^{(1)}$ and $H^{(2)}$ of the specified TPS $\mathcal{C}$. Note that under the above product unitary operations the local coherence within individual subsystem has been eliminated while the quantum correlation is invariant.

Since quantum correlation is a kind of nonlocal coherence, it should be invariable under any local unitary operation, and a proper measure of nonlocal coherence is supposed to measure quantum correlation as well. Mathematically, the off-diagonal elements of the density matrix is often employed to measure the overall coherence of the system, including the local coherence within individual subsystems. This lead to a direct conjecture that after eliminating the local coherence by local unitary operations, the transformed off-diagonal elements of the density matrix should only characterize the total nonlocal coherence, namely all the quantum correlation. For pure states the task is convenient to be done by Schmidt decomposition mentioned above, for bipartite mixed states the task is to reduce the local coherence to zero by local unitary operations. Then the quantum correlation can be measured as the following way.

A measure of bipartite quantum correlation for a system with TPS $\mathcal{C}$, we call it consonance $\mathfrak{C}$, is defined as
\begin{eqnarray}
\mathfrak{C}(\rho)\coloneqq\inf\left\{\sum_{ijmn}\left|\rho_{ij,mn}^{c}\left(1-\delta_{im}\right)\left(1-\delta_{jn}\right)\right|\right\},\label{mc}
\end{eqnarray}
where $\rho^{c}=\left(U_{1}\otimes U_{2}\right)\rho\left(U_{1}\otimes U_{2}\right)^{\dag}$ denotes the state $\rho$ underwent any local unitary operation to attain the vanishing of local coherence
\begin{eqnarray}
\mathcal{L}(\rho)\coloneqq\sum_{i=m,j\neq{n}}\left|\rho_{ij,mn}^c\right|+\sum_{i\neq{m},j=n}\left|\rho_{ij,mn}^c\right|=0.\label{lc}
\end{eqnarray}
Here $i,m$ and $j,n$ label the orders of basis with Hilbert spaces $H^{(1)}$ and $H^{(2)}$, respectively. In subsequent paragraphs we study the relation between consonance, quantum discord, and entanglement for various initial states.

\textit{1. Bell-like states.} For pure states, to seek for the infimum in Eq.~(\ref{lc}), it is equivalent to perform the Schmidt decomposition on the initial state. As the first example, we take Bell states $\ket{\Psi^\pm}$ and $\ket{\Phi^\pm}$. It easily can be seen that for this particular case the consonance and any other measure of quantum correlation coincide, which means, $\mathfrak{C}(\rho)=\mathcal{D}(\rho)=C(\rho)=1$. Then we consider the Bell-like states $\ket{\phi}=a\ket{11}_{12}+b\ket{00}_{12}$ and $\ket{\psi}=a\ket{10}_{12}+b\ket{01}_{12}$ $(|a|^2+|b|^2=1)$, which are already in the Schmidt decomposed form and have a simple analytical expression of consonance. The consonance is just equal to the concurrence, that is, $\mathfrak{C}\left(\rho\right)=C\left(\rho\right)=2|ab|$ , while the discord is equal to the entanglement of formation (EoF) \cite{Wooters98,Bennett96},
\begin{eqnarray}
\mathcal{D}(\rho)=&&\mathcal{E}(\rho)\nonumber\\
=&&1-(1/2)\left[\left(1+\sqrt{1-4|ab|^2}\right)\log_2\left(1+\sqrt{1-4|ab|^2}\right)\right.\nonumber\\
&&\left.+\left(1-\sqrt{1-4|ab|^2}\right)\log_2\left(1-\sqrt{1-4|ab|^2}\right)\right].
\end{eqnarray}
We note that the consonance is larger than quantum discord except for the trivial cases (separable or maximally entangled). Besides, the monotonicity of consonance is essentially the same as that of quantum discord.

\textit{2. General $2\otimes2$ pure states.} In fact for any $2\otimes2$ pure states, the consonance is equal to the concurrence. After Schmidt decomposing on the general pure state $\ket{\varphi}=a\ket{1}_1\ket{1}_{2}+b\ket{1}_1\ket{0}_{2}+c\ket{0}_1\ket{1}_{2}+d\ket{0}_1\ket{0}_{2}$, one can obtain $\ket{\varphi}=\sqrt{\lambda_1}\ket{\tilde{0}\tilde{0}}_{12}+\sqrt{\lambda_2}\ket{\tilde{1}\tilde{1}}_{12}$, where $\lambda_1$ and $\lambda_2$ are the eigenvalues of either reduced density matrix. Thus by the definition in Eq.~(\ref{mc}), the consonance is $\mathfrak{C}(\rho)=2\sqrt{\lambda_1\lambda_2}$. Through simple calculation the eigenvalues are obtained as $\lambda_1=(1/2)[1+\sqrt{1-4(ad-bc)^2}]$ and $\lambda_2=(1/2)[1-\sqrt{1-4(ad-bc)^2}]$. Therefore $\mathfrak{C}(\rho)=2|ad-bc|$, which is just equal to the concurrence \cite{Hill97}.

\begin{figure}
\includegraphics[width=7cm]{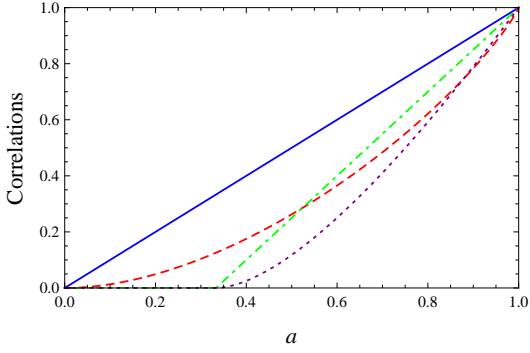}
\caption{\label{fig:2}(Color online) Graphs of consonance $\mathfrak{C}(\rho)$ (solid, blue), quantum discord $\mathcal{D}(\rho)$ (dashed, red), concurrence $C(\rho)$ (dotted-dashed, green), and entanglement of formation $\mathcal{E}(\rho)$ (dotted, purple) versus $a$ for the Werner state $\rho=a\ket{\Psi^-}\bra{\Psi^-}+(1-a)I/4$.}
\end{figure}

\textit{3. The Werner state.} It is convenient to study the bipartite mixed states with $X$ type density matrix \cite{Yu07}, for which the local coherence within individual subsystem is naturally vanishing thus local unitary operations are not required to obtain the infimum in Eq.~(\ref{lc}). Let us consider the Werner state. Besides the expressions of concurrence and discord mentioned above, we have the consonance $\mathfrak{C}(\rho)=a$, and EoF written in terms of concurrence $\mathcal{E}(\rho)=-f(\rho)\log_2f(\rho)-[1-f(\rho)]\log_2[1-f(\rho)]$, where $f(\rho)=[1+\sqrt{1-C(\rho)^2}]/2$ \cite{Wooters98}. The graphs of $\mathfrak{C}(\rho)$, $\mathcal{D}(\rho)$, $C(\rho)$, and $\mathcal{E}(\rho)$ versus $a\in\left[0,1\right]$ are depicted in Fig.~(\ref{fig:2}). In contrast to the different orders of quantum discord, concurrence, and EoF as functions of $a$ \cite{Luo08b,Ali10a}, the consonance is always larger than the other three correlations except for the trivial cases, and the consonance and quantum discord still have the same monotonicity.

\begin{figure}
\includegraphics[width=7cm]{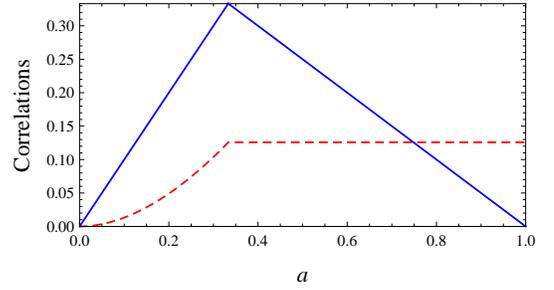}
\caption{\label{fig:3}(Color online) Graphs of consonance minus concurrence $\mathfrak{C}(\rho)-C(\rho)$ (solid, blue) and quantum dissonance $\mathcal{Q}(\rho)$ (dashed, red) versus $a$ for the Werner state.}
\end{figure}

The above examples show the quantitatively and qualitatively similarity between the consonance and concurrence, and indicate the rest of quantum correlation excluding entanglement can be quantified by subtracting concurrence from consonance, which is plotted in Fig.~(\ref{fig:3}) in comparison with the quantum dissonance $\mathcal{Q}(\rho)$ \cite{Modi10}. The two measures attain maximum simultaneously, but the dissonance is invariable for any entangled Werner state (broadly for any entangled Bell-diagonal state), resulting from the fixedness of the closest separable state $\sigma$ and classical state $\chi_\rho$. Meanwhile the consonance minus concurrence decreases subsequently and vanishes for the maximum entangled state. The diverse behaviors do not imply contradictory but reflect the qualitatively discrepancy between these two notions of correlation, that is, the consonance minus concurrence quantify the non-entangled component within quantum correlation of a given state, while the dissonance quantify the nonclassical correlation with the absence of entanglement.

\begin{figure}
\includegraphics[width=7cm]{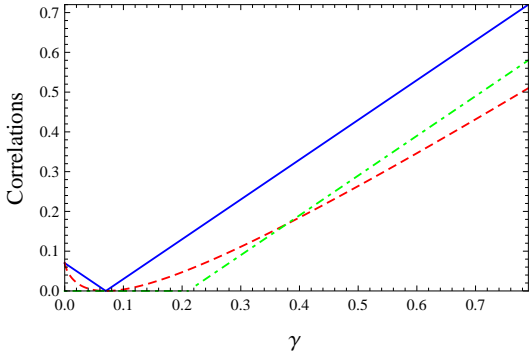}
\caption{\label{fig:4}(Color online) Graphs of consonance $\mathfrak{C}(\rho)$ (solid, blue), quantum discord $\mathcal{D}(\rho)$ (dashed, red), and negativity $\mathcal{N}(\rho)$ (dotted-dashed, green) versus $\gamma$ for the class of states in Eq.~(\ref{2x3}), with the parameter chosen as $\beta=0.07$. The graphs show that for $\beta=\gamma$, all correlations are equal to zero.}
\end{figure}

\textit{4. The two-parameter class of states with $2\otimes3$ Hilbert space}. Finally we consider the states describing a qubit-qutrit quantum system with the TPS $C=H^{(1)}\otimes H^{(2)}$, where $H^{(1)}$ is 2-dimension and $H^{(2)}$ is 3-dimension, respectively. The class of states with two real parameters $\alpha$ and $\gamma$ takes the form as \cite{Ali10b}
\begin{eqnarray}
\rho=&&\alpha\left(\ket{02}\bra{02}+\ket{12}\bra{12}\right)+\beta\left(\ket{\Phi^+}\bra{\Phi^+}\right.\nonumber\\
&&\left.+\ket{\Phi^-}\bra{\Phi^-}+\ket{\Psi^+}\bra{\Psi^+}\right)
+\gamma\ket{\Psi^-}\bra{\Psi^-},\label{2x3}
\end{eqnarray}
where parameter $\beta$ depends on $\alpha$ and $\gamma$ by the unit trace condition $2\alpha+3\beta+\gamma=1$, and the ranges of parameters are $0\leq a\leq 1/2$ and $0\leq\gamma\leq1$. For these states the local coherence is also vanishing. We adopt the negativity to measure the entanglement of $\rho$, which is given as \cite{Chi03}
$\mathcal{N}(\rho)=\max\left\{0,2\alpha+2\gamma-1\right\}$.
The quantum discord of $\rho$, recently derived by Ali \cite{Ali10b}, is given by
\begin{eqnarray}
\mathcal{D}(\rho)=\beta\log_2(2\beta)+\gamma\log_2(2\gamma)-\left(\beta+\gamma\right)\log_2\left(\beta+\gamma\right),
\end{eqnarray}
and we can simply obtain the consonance as $\mathfrak{C}(\rho)=\left|\beta-\gamma\right|$. We note that with $\gamma=0$, the negativity is vanishing, while the consonance and quantum discord coincide, that is, $\mathfrak{C}(\rho)=\mathcal{D}(\rho)=(1-2\alpha)/3$. With $\beta=0$, all three quantum correlations coincide, that is, $\mathfrak{C}(\rho)=\mathcal{D}(\rho)=\mathcal{N}(\rho)=1-2\alpha$. With $\alpha,\beta,\gamma>0$, the consonance, quantum discord, and negativity versus $\gamma$ are displayed in Fig.~(\ref{fig:4}) for a specific value of the parameter $\beta=0.07$. For this particular initial state, the dominance relation among correlations still exists, being similar with the case for Werner state. The monotonicity of consonance and quantum discord are still the same. For $\beta=\gamma$, $\rho$ becomes diagonal and indicate a classical state without any quantum correlation, thus the consonance, quantum discord, and negativity all vanishes.

For multipartite systems, we do not want to go into too much detail here but give the generalized form of the consonance for arbitrary dimension multipartite quantum states. The TPS of the total Hilbert space is $\mathcal{C}=\otimes_{i=1}^n{H^{(i)}}$, where $H^{(i)}$ is the $i$-th $d_i$-dimension Hilbert space with the basis $\{\ket{k}_i|k=1,2,\cdots,d_i\}$. Thus the consonance for $n$-partite system takes the form
\begin{eqnarray}
\mathfrak{C}(\rho)\coloneqq\inf\left\{\sum_{i_1i_2\cdots{i_n},j_1j_2\cdots{j_n}}\left|\rho_{i_1i_2\cdots{i_n},j_1j_2\cdots{j_n}}^{c}\prod_k\left(1-\delta_{i_kj_k}\right)\right|\right\},
\end{eqnarray}
where $\rho^{c}=U\rho{U}^{\dag}$ denotes the state $\rho$ underwent any unitary operation (excluding global unitary operation) to attain the infimum of local coherence
\begin{eqnarray}
\mathcal{L}(\rho)\coloneqq\inf\left\{\sum_{i_1i_2\cdots{i_n},j_1j_2\cdots{j_n}}\left|\rho_{i_1i_2\cdots{i_n},j_1j_2\cdots{j_n}}^cf\left(i_1\cdots{i_n},j_1\cdots{j_n}\right)\right|\right\}.
\end{eqnarray}
Here the function $f$ excludes the diagonal elements and elements with completely different corresponding subscripts and it takes the form
\begin{eqnarray}
f\left(i_1\cdots{i_n},j_1\cdots{j_n}\right)=\left(1-\prod_k{\delta_{i_kj_k}}\right)\left[1-\prod_k\left(1-\delta_{i_kj_k}\right)\right].
\end{eqnarray}
This generalized form for tripartite system can easily distinguish \textrm{GHZ} state and $W$ state \cite{Coffman00}, that is, $\mathfrak{C}(\ket{\textrm{GHZ}}\bra{\textrm{GHZ}})=0$ and $\mathfrak{C}(\ket{W}\bra{W})=1$, as well as the 3-tangle $\tau$ can \cite{Dur00}.

We have reexamined quantum correlation from the view of its consanguineous quantum property, the coherence, and quantify quantum correlation by nonlocal coherence. The importance of specifying the TPS of the total state space is emphasized. We have demonstrated there are clear dominance relations between the consonance and other measures of quantum correlation, and the consonance presents the same monotonicity as that of quantum discord. The qualitatively similarity between consonance and concurrence implies a meaningful way to quantify the rest of quantum correlation excluding entanglement. Our perspective has rather fundamental and general meaning, thus the measure can be easily generalized to the multipartite case. We hope our methods will quantify genuine multipartite quantum correlation felicitously.



Furthermore, in specific physical systems, it is not an easy task to find the direct relation between some mechanical quantity and quantum correlation, e.g. quantum discord. Since our measure relates the quantum correlation and coherence closely, it not only provides theory evidence to experimentally measure the entanglement by some quantity e.g. the visibility of the interference \cite{Pathak10} and anti-diagonal density matrix elements from polarization tomography measurements \cite{Akopian06}, but also suggests the possibility to construct measurement scheme for quantum correlation in a direct way. Namely to measure quantum correlation, one need not the full knowledge of the system, but only to find some measurable quantity reflecting the nonlocal coherence between subsystems. Meanwhile, our measure gives a limitation for this application of correlation measurement, which means the local coherence should be vanishing, otherwise the results will not faithfully reflect the nonlocal coherence. For states with the presence of local coherence, the reprocessing of measurement results to obtain the infimum may be still required.



This work is supported by the National Nature Science Foundation of China under Grants No. 60703100 and No. 10775023, and by the Fundamental Research Funds for the Central Universities under Grant No. DUT10LK10.


\begin{thebibliography}{5}
\bibitem{Einstein35}A. Einstein, B. Podolosky, and N. Rosen, Phys. Rev. \textbf{47} 777 (1935);
\bibitem{Schrodinger35}E. Schr\"{o}dinger, Naturwiss. \textbf{23}, 807 (1935).

\bibitem{Horodecki09}R. Horodecki, P. Horodecki, M. Horodecki, and K. Horodecki, Rev. Mod. Phys. \textbf{81}, 865 (2009).

\bibitem{Werner89}R. F. Werner, Phys. Rev. A \textbf{40}, 4277 (1989).


\bibitem{Bennett99}C. H. Bennett, D. P. DiVincenzo, C. A. Fuchs, T. Mor, E. Rains, P. W. Shor, J. A. Smolin, and W. K. Wootters, Phys. Rev. A \textbf{59}, 1070 (1999).
\bibitem{Horodecki05}M. Horodecki, P. Horodecki, R. Horodecki, J. Oppenheim, A. Sen, U. Sen, and B. Synak-Radtke, Phys. Rev. A \textbf{71}, 062307 (2005).
\bibitem{Niset06}J. Niset and N. J. Cerf, Phys. Rev. A \textbf{74}, 052103 (2006).

\bibitem{Braunstein99}S. L. Braunstein, C. M. Caves, R. Jozsa, N. Linden, S. Popescu, and R. Schack, Phys. Rev. Lett. \textbf{83}, 1054 (1999).
\bibitem{Meyer00}D. A. Meyer, Phys. Rev. Lett. \textbf{85}, 2014 (2000).
\bibitem{Biham04}E. Biham, G. Brassard, D. Kenigsberg, and T. Mor, Theor. Comput. Sci. \textbf{320}, 15 (2004).
\bibitem{Datta05}A. Datta, S. T. Flammia, and C. M. Caves, Phys. Rev. A \textbf{72}, 042316 (2005);
\bibitem{Datta07}A. Datta and G. Vidal, ibid. \textbf{75}, 042310 (2007).

\bibitem{Henderson01}L. Henderson and V. Vedral, J. Phys. A \textbf{34}, 6899 (2001).
\bibitem{Ollivier01}H. Ollivier and W. H. Zurek, Phys. Rev. Lett. \textbf{88}, 017901 (2001).
\bibitem{Oppenheim02}J. Oppenheim, M. Horodecki, P. Horodecki, and R. Horodecki, Phys. Rev. Lett. \textbf{89}, 180402 (2002).
\bibitem{Groisman05}B. Groisman, S. Popescu, and A. Winter, Phys. Rev. A \textbf{72}, 032317 (2005).
\bibitem{Luo08a}S. Luo, Phys. Rev. A \textbf{77}, 022301 (2008).


\bibitem{Luo08b}S. Luo, Phys. Rev. A \textbf{77}, 042303 (2008).

\bibitem{Modi10}K. Modi, T. Paterek, W. Son, V. Vedral, and M. Williamson, Phys. Rev. Lett. \textbf{104}, 080501 (2010).

\bibitem{Yu09}C. -S. Yu and H. -S. Song, Phys. Rev. A \textbf{80}, 022324 (2009).

\bibitem{Zanardi04}P. Zanardi, D. Lidar, and S. Lloyd, Phys. Rev. Lett. \textbf{92}, 060402 (2004).

\bibitem{Li07}C. Li, H. -S. Song, and L. Zhou, Int. J. Theor.
Phys. \textbf{46} 1815 (2007).

\bibitem{Wooters98}W. K. Wootters, Phys. Rev. Lett. \textbf{80}, 2245 (1998).

\bibitem{Ali10a}M. Ali, A. R. P. Rau, and G. Alber, Phys. Rev. A \textbf{81}, 042105 (2010).


\bibitem{Bennett96}C. H. Bennett, D. P. DiVincenzo, J. A. Smolin, and W. K. Wootters, Phys. Rev. A \textbf{54}, 3824 (1996).

\bibitem{Hill97}S. Hill and W. K. Wootters, Phys. Rev. Lett. \textbf{78}, 5022 (1997).

\bibitem{Yu07}T. Yu and J. H. Eberly, Quantum Inform. Comput. \textbf{7}, 459 (2007).

\bibitem{Ali10b}M. Ali, J. Phys. A: Math. Theor. \textbf{43}, 495303 (2010).


\bibitem{Chi03}D. P. Chi and S. Lee, J. Phys. A: Math. Gen. \textbf{36}, 11503 (2003).

\bibitem{Pathak10}P. K. Pathak and S. Hughes, arXiv:1010.1713.

\bibitem{Akopian06}N. Akopian, N. H. Lindner, E. Poem, Y. Berlatzky, J. Avron, and D. Gershoni, Phys. Rev. Lett. \textbf{96}, 130501 (2006).

\bibitem{Coffman00}V. Coffman, J. Kundu, and W. K. Wootters, Phys. Rev. A \textbf{61}, 052306 (2000).

\bibitem{Dur00}W. D\"{u}r, G. Vidal, and J. I. Cirac, Phys. Rev. A \textbf{62} 062314 (2000).








\end{thebibliography}
\end{document}